\documentclass[%
 reprint,
superscriptaddress,
nofootinbib,
 amsmath,amssymb,
 aps,
]{revtex4-1}
\usepackage{graphics,epsfig,psfrag}
\usepackage{graphicx}
\usepackage{dcolumn}
\usepackage{color}
\usepackage{bm}

\begin{document}

\preprint{APS/123-QED}

\title{Analytical calculation of black hole spin using deformation of the shadow}

\author{Oleg Yu. Tsupko}
\email{tsupko@iki.rssi.ru}
\affiliation{Space Research Institute of Russian Academy of Sciences, Profsoyuznaya 84/32, Moscow 117997, Russia}
\affiliation{National Research Nuclear University MEPhI (Moscow Engineering Physics Institute),\\ Kashirskoe Shosse 31, Moscow 115409, Russia}%

\date{\today}

\begin{abstract}
We succeed to find compact analytical expressions which allow to easily extract the black hole spin from observations of its shadow, without need to construct or model the entire curve of the shadow. The deformation of Kerr black hole shadow can be characterized in a simple way by oblateness (the ratio of the horizontal and vertical angular diameters which are supposed to be measured by an observer). The deformation is significant in case the black hole is nearly extreme and observer is not so far from the equatorial plane. In this approximation, we present: (i) the spin lower limit via oblateness, (ii) the spin via oblateness and viewing angle, in case the latter is known from other observations.



\end{abstract}

\pacs{?????? - ??????}
\maketitle


\section{Introduction}

For a distant observer, a black hole (BH) should be seen as a dark spot in the sky which is referred to as a 'BH shadow'. More rigorously, the shadow can be defined as the region of the observer's sky that is left dark if there are light sources distributed everywhere but not between the observer and the BH \cite{GrenzebachPerlickLaemmerzahl2014}. Size and shape of the shadow are determined by parameters of the BH and the observer position. At present, an increasing interest concerning investigations of the shadow is connected with the challenging perspective of possible observation of the shadow of the supermassive BH in the center of our Galaxy. Two projects are under way now to observe this shadow: the Event Horizon
Telescope (http://eventhorizontelescope.org) and the BlackHoleCam (http://blackholecam.org).

Using estimates of the BH mass, we can calculate the assumed size of its shadow (if the BH-observer distance is also known). \textit{Vice versa}, what we can get from observation of the shadow angular radius is the BH mass.

\begin{figure}
	\centerline{\hbox{\includegraphics[width=0.30\textwidth]{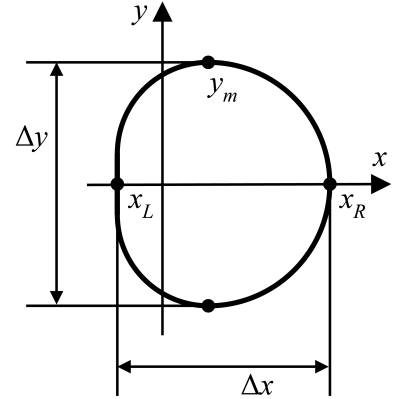}}}
	\caption{The simplest way to characterize the deformation of the shadow is to use oblateness, the ratio of horizontal ($\Delta x$) and vertical ($\Delta y$) diameters of the shadow which are supposed to be measured by an observer ($\Delta x \le \Delta y$). The oblateness $k = \Delta x / \Delta y$ ranges from $1$ (Schwarzschild, no deformation) to $\sqrt{3}/2$ (extreme Kerr, the strongest deformation). For analytical calculation of diameters, we need to know the left and the right horizontal borders of the shadow, $x_L$ and $x_R$, and the vertical border, $y_m$.}
	\label{figure-oblateness}
\end{figure}

If a BH is rotating, the shadow is not circular, but oblate and deformed. The second thing we could hope to measure is the oblateness (the ratio of the horizontal and vertical angular diameters) of the shadow, see Fig. \ref{figure-oblateness}. The oblateness can give us information about the BH spin. It is important also that the deformation depends on the viewing angle of observer: for the equatorial observer the deformation is strongest, while for the polar observer the deformation is absent.

\begin{figure*}
	\centerline{\hbox{\includegraphics[width=0.9\textwidth]{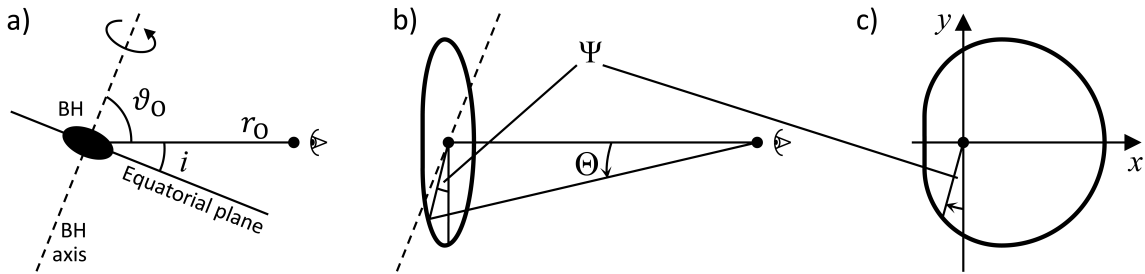}}}
	\caption{Geometry of the problem. a) Position of the observer and the black hole. We assume that $r_{\textrm{O}} \gg m$. We denote $\vartheta_{\textrm{O}}$ the inclination angle, and $i$ the viewing angle. Main results are obtained for the nearly equatorial observer, which means that $i \ll 1$. b) and c) Celestial coordinates of the observer. $\Theta$ is the colatitude, and $\Psi$ is the azimuthal angle, see also \cite{GrenzebachPerlickLaemmerzahl2014} and \cite{GrenzebachPerlickLaemmerzahl2015}. $x$ and $y$ are Cartesian coordinates calculated for $r_{\textrm{O}} \gg m$ by the formulas (\ref{Cartesian}). Note that in our coordinates, the origin corresponds to a principal null ray, and in figures in \cite{Bardeen1973} the origin is determined by zero impact parameters. Therefore in our case the origin is horizontally shifted by the value $a \sin \vartheta_{\textrm{O}}$ in comparison with \cite{Bardeen1973}. For example, for the extreme Kerr BH, the shadow is situated between $-m/r_{\textrm{O}}$ and $8m/r_{\textrm{O}}$ in our paper, whereas in \cite{Bardeen1973} it lies between $-2m$ and $7m$. For details see \cite{PerlickTsupko2017}.}
	\label{figure-observer}
\end{figure*}

Shadow was extensively studied in literature, which includes analytical investigations and numerical simulations (for example, see \cite{Synge1966, Bardeen1973, GrenzebachPerlickLaemmerzahl2014, GrenzebachPerlickLaemmerzahl2015,  FalckeMeliaAgol2000, FrolovZelnikov2011, JamesTunzelmannFranklinThorne2015, Luminet1979, Chandra, Konoplya2016, Tsupko2015, Johannsen2016,  HiokiMaeda2009, LiBambi2014, TsukamotoLiBambi2014, Takahashi2004, YangLi2016, Abdujabbarov2015, Bambi2013, Dymnikova1986, Zakharov2014, PerlickTsupko2017}).
Extraction of the spin from the shadow deformation was discussed in number of papers \cite{HiokiMaeda2009, LiBambi2014, TsukamotoLiBambi2014, Takahashi2004, YangLi2016, Abdujabbarov2015, Bambi2013}, starting from work of Hioki and Maeda \cite{HiokiMaeda2009} with distortion parameter. These works imply the use of numerical calculations at some stage, and to the best of our knowledge, there is no fully analytical treatment of the problem. We believe that an explicit analytical dependence of the spin on some parameter characterizing the shadow non-sphericity and observer viewing angle would be very useful as a first step in the development of more complex models.

Analytical investigations of the BH shadow start from work of Synge \cite{Synge1966}, where the angular radius of the shadow was calculated for the Schwarzschild BH, as a function of the BH mass and of the radial coordinate of the observer. The shape of the Kerr BH shadow was calculated by Bardeen \cite{Bardeen1973}. In the paper \cite{GrenzebachPerlickLaemmerzahl2014}, the size and the shape of the shadow were calculated for the whole class of Pleba{\'n}ski-Demia{\'n}ski spacetimes.

Results of the paper \cite{GrenzebachPerlickLaemmerzahl2014} allow anyone to calculate the shadow of Kerr BH for any position of the observer, which means arbitrary radial distance from BH and arbitrary inclination of observer. Nevertheless, \textit{analytical} calculation of the horizontal and vertical angular diameters in general case is complicated. Analytical calculation of the shadow means the following: every point of the curve is evaluated as an analytical function of a special parameter, see details below. This parameter is changed in some range, and boundaries of this range are also subject of evaluation. Namely, we need to find zeros of a high-order polynomials. Therefore in the general case results for diameters can not be presented in closed analytical form (as explicit functions of spin and inclination). Calculation of the horizontal and vertical angular diameters is addressed in the paper \cite{GrenzebachPerlickLaemmerzahl2015}. The authors consider the equatorial plane of the Kerr BH and explain how to calculate the horizontal and vertical angular diameters of the shadow as a function of the BH mass, spin, and the radial coordinate of the observer. As an example of the situation when results can be written explicitly, the authors have calculated the horizontal and vertical angular diameters of the shadow for extreme Kerr BH.

Our goal is to obtain a simple analytical dependence of oblateness on the spin and inclination which will be easy to use. This goal is achieved by using the approximation of a nearly extreme BH with $a=(1-\delta)m$, $\delta \ll 1$. In this approximation, it becomes possible to obtain an explicit dependence which, however, is still too cumbersome. For further simplifications, we consider the case of nearly equatorial observer.

Remarkably, we obtain that the dependence of the deformation on the spin is strong: the oblateness is proportional to $\sqrt{\delta}$. It means that a small deviation of spin from the extreme value leads to a notable change of the shadow. At the same time, the dependence of deformation on the viewing angle is quadratic and therefore not as important. 

In practical situations, it is expected that the observer could measure the horizontal and vertical angular diameters of the BH, and knows the oblateness. Therefore we reformulate our results as a technique of extraction of spin by measuring the oblateness.


As a main result, we present compact formulas for: (i) expression of the spin lower limit via oblateness; (ii) direct calculation of the spin via oblateness and viewing angle, in case the latter is known from other observations.


The paper is organized as follows. In the next Section we explain how to calculate the shadow in  case of arbitrary observer's inclination angle. In Section \ref{sec: nearly} we find explicit dependence of spin on the oblateness and observer's viewing angle, for the case the black hole is nearly extreme and observer is near equatorial plane; then we come to Conclusions.

\section{Construction of the shadow and extraction of BH spin for arbitrary observer's inclination angle}

We will work in the Kerr metric with $G=c=1$:
\begin{gather}\label{eq:g}
ds^2 =  
 - \left(1-\frac{2mr}{\rho^2}\right) dt^2 + 
\frac{\rho^2}{\Delta} dr^2 + \rho^2 d\vartheta^2 
\nonumber
\\
+ \, \mathrm{sin}^2\vartheta\left(r^2+a^2+
\frac{2mra^2\mathrm{sin}^2\vartheta}{\rho^2}\right)
d\varphi^2
\nonumber
\\
- \frac{4mra \mathrm{sin} ^2\vartheta}{\rho^2} \, dt \, d\varphi
\end{gather}
where
\begin{equation}\label{eq:Deltarho}
\Delta = r^2 + a^2 - 2mr \, , \quad
\rho ^2 = r^2 + a^2 \mathrm{cos} ^2 \vartheta \, .
\end{equation}

We consider an observer at the position $(r_{\textrm{O}}, \vartheta_{\textrm{O}})$. Equations for calculation of the shadow curve for this observer can be found from the equations (24)--(26) of Grenzebach, Perlick, L{\"a}mmerzahl \cite{GrenzebachPerlickLaemmerzahl2014} simplified for the Kerr metric:
\begin{equation} \label{Theta}
\sin \Theta(r) = \frac{2r\sqrt{r^2+a^2-2mr} \sqrt{r_{\textrm{O}}^2+a^2-2m r_{\textrm{O}}}}{ r_{\textrm{O}}^2 r - r_{\textrm{O}}^2 m + r^3 -3r^2 m + 2ra^2} \, ,
\end{equation} 
\begin{equation}
\sin \Psi(r) = - \frac{r^3 - 3r^2m + ra^2 + a^2 m + a^2 \sin^2 \vartheta_{\textrm{O}} (r-m)}{2ar \sin \vartheta_{\textrm{O}} \sqrt{r^2+a^2-2mr}} \, ,
\end{equation}
where $\Theta$ and $\Psi$ are the celestial coordinates for our observer, see Fig. \ref{figure-observer}. These two angles determine the shape of the shadow as a function of the parameter $r$ which means the radius of critical spherical photon orbit. Parameter $r$ is changed from its minimal $r_{min}$ to maximum $r_{max}$ value, they are found from Eqs
\begin{equation} \label{min}
\sin \Psi(r) = 1 \text{  for  } r_{min}, \text{  and}
\end{equation}
\begin{equation} \label{max}
\sin \Psi(r) = - 1 \text{  for  } r_{max} \, .
\end{equation}

%

We restrict ourselves to the consideration of distant observer, and for $r_{\textrm{O}} \gg m$, the formula (\ref{Theta}) can be simplified:
\begin{equation} \label{Theta-simple}
\sin \Theta(r) = \frac{2r\sqrt{r^2+a^2-2mr}}{r_{\textrm{O}}(r-m)} \, .
\end{equation}

It is convenient to use dimensionless Cartesian coordinates in observer's sky (see \cite{GrenzebachPerlickLaemmerzahl2014}):
\begin{equation}\label{eq:stereo}
x = -2 \tan ( \Theta/2 )
		\sin \Psi  \, , \quad y = -2 \tan ( \Theta/2 )
		\cos \Psi .
\end{equation}
For $r_{\textrm{O}} \gg m$ angular size of the shadow is very small, $\Theta \ll 1$, and in this approximation we can write for convenience that
\begin{equation} \label{Cartesian}
x = -  \sin \Theta \sin \Psi \, , \quad
y = -  \sin \Theta \cos \Psi \, .
\end{equation}

The shape of the shadow can be characterized by its left border $x_L < 0$, the right border $x_R>0$, and the maximum value of $y$-coordinate, $y_m$, see Fig. \ref{figure-oblateness}. These values give us the 'horizontal' $\Delta x = x_R - x_L$ and 'vertical' $\Delta y = 2y_m$ diameters of the shadow. We are interested in calculation of oblateness $k = \Delta x / \Delta y$.

\begin{figure*}
	\centerline{\hbox{\includegraphics[width=0.8\textwidth]{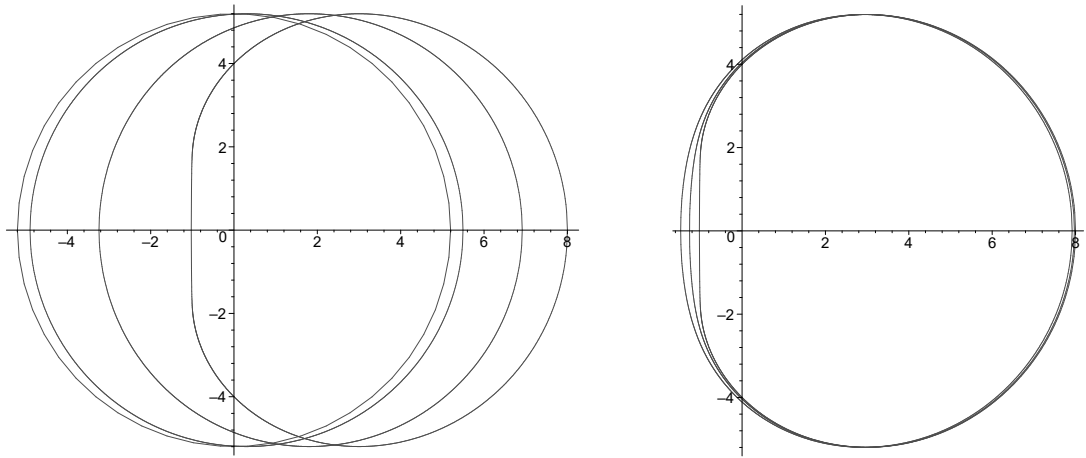}}}
	\caption{LEFT: The shadow curves for the distant equatorial observer for (from the leftmost to the rightmost) $a=0$, $0.1 m$, $0.6 m$, $0.9999 m$. RIGHT: The shadow curves for $a=0.97m$, $0.99m$, $0.9999m$. There is a notable difference in location of left borders, whereas the right borders are approximately at the same place, see (\ref{x_L_delta}) and (\ref{x_R_delta}).}
	\label{figure-fig3}
\end{figure*}

\begin{figure*}
	\centerline{\hbox{\includegraphics[width=0.8\textwidth]{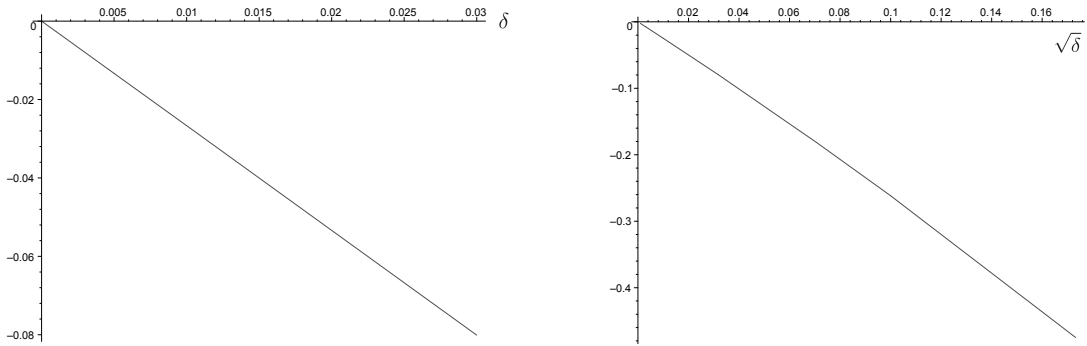}}}
	\caption{LEFT: The left hand side of (\ref{x_R_delta}) is plotted numerically as a function of $\delta$. Change in position of right border of the shadow is small. RIGHT: The left hand side of (\ref{x_L_delta}) is plotted numerically as a function of $\sqrt{\delta}$, at the same range of $\delta$ as on the left figure. There is a significant change in position of left border of the shadow corresponding to small change of $\delta$.}
	\label{figure-PRDris02}
\end{figure*}

Let us consider the equatorial observer. For the Schwarzschild case, the shadow is circular and $|x_L| = x_R = y_m = 3\sqrt{3} \, m / r_{\textrm{O}}$. With increasing of $a$, the shadow is shifted to the right, see Fig. 3 (left). At small $a \ll m$, the left and right borders are shifted equally, and the horizontal diameter $\Delta x = 6\sqrt{3} \, m / r_{\textrm{O}}$ is not changing \cite{FrolovZelnikov2011}. At $a=m$, the borders tend to $x_L=-m/ r_{\textrm{O}}$ and $x_R = 8m/ r_{\textrm{O}}$. At the same time, the vertical diameter stays constant $\Delta y = 6 \sqrt{3} m / r_{\textrm{O}}$ for all values of $a$ \cite{GrenzebachPerlickLaemmerzahl2015}. Therefore, for the equatorial observer, the oblateness ranges from $1$ for the Schwarzschild case ($a=0$) to $\sqrt{3}/2$ for the extreme Kerr case ($a=m$).

Let us now consider the nearly extreme Kerr BH $a=(1-\delta) m$ with $\delta \ll 1$. The remarkable thing we have seen from plotting the shadow is that the displacement of the left border in comparison with the extreme Kerr case is proportional to $\sqrt{\delta}$:
\begin{equation} \label{x_L_delta}
x_L|_{a=(1-\delta)m} - x_L|_{a=m} \, \propto \, \sqrt{\delta} \, ,
\end{equation}
whereas the right border is shifting proportionally to $\delta$:
\begin{equation} \label{x_R_delta}
x_R|_{a=(1-\delta)m} - x_R|_{a=m}  \, \propto  \, \delta \, , \;\; \text{see FIG. 3 (right)} \, .
\end{equation}
Left hand sides of these formulae are shown in FIG. \ref{figure-PRDris02}.

With this in mind, we seek the solution of (\ref{min}) in the form
\begin{equation}
r_{min} (\delta, \vartheta_{\textrm{O}}) = r_0(\vartheta_{\textrm{O}}) + r_1(\vartheta_{\textrm{O}}) \sqrt{\delta} + r_2(\vartheta_{\textrm{O}}) \delta \, , \;\; \delta \ll 1 \, .
\end{equation}
Substituting the expressions for $a$ and $r_{min}$ in the equation (\ref{min}) and keeping the terms with $\sqrt{\delta}$ and $\delta$, we obtain the three equations for the unknowns $r_0(\vartheta_{\textrm{O}})$, $r_1(\vartheta_{\textrm{O}})$, $r_2(\vartheta_{\textrm{O}})$. The equation for $r_0(\vartheta_{\textrm{O}})$ is polinomial and has several solutions. We need to choose the one which is $r_0(\vartheta_{\textrm{O}}) \gtrsim m$ and tends to $m$ when $\vartheta_{\textrm{O}} \rightarrow \pi/2$. We get that $r_0(\vartheta_{\textrm{O}}) = m$ \footnote{It should be noted that this is true only for $\arcsin(\sqrt{3}-1) \le \vartheta_{\textrm{O}} \le \pi/2$ (see, for example, \cite{Dymnikova1986}), but it implies $i < 43^\circ$, so it is enough for our purposes.}, and then find $r_1(\vartheta_{\textrm{O}})$ and $r_2(\vartheta_{\textrm{O}})$.
The left border of the shadow is calculated as
\begin{equation}
x_L (\delta, \vartheta_{\textrm{O}}) = - \sin \Theta(r_{min}) \, .
\end{equation}
In this manner, we find $x_L$ up to the terms $\propto \sqrt{\delta}$. The terms $\propto \sqrt{\delta}$ are presented only in $x_L$, $\delta$-corrections in all other expressions start from $\delta$ and therefore can be neglected. It means that in all other values we can put $a=m$.

Solving (\ref{max}) for $r_{max}$ with $a=m$, we choose a root which tends to $4m$ when $\vartheta_{\textrm{O}} \rightarrow \pi/2$:
\begin{equation}
r_{max} (\vartheta_{\textrm{O}}) =  \left(\sin \vartheta_{\textrm{O}} +1 + \sqrt{ 2 \sin \vartheta_{\textrm{O}} + 2 } \right) m \, .
\end{equation}
The right border of the shadow is calculated as
\begin{equation}
x_R (\delta, \vartheta_{\textrm{O}}) = \sin \Theta(r_{max}) \, .
\end{equation}
Horizontal size of the shadow has the form:
\begin{equation}
\Delta x = x_R- x_L = F_0(\vartheta_{\textrm{O}}) + F_1(\vartheta_{\textrm{O}}) \sqrt{\delta} \, ,
\end{equation}
where $F_0(\vartheta_{\textrm{O}})$ and $F_1(\vartheta_{\textrm{O}})$ are functions too cumbersome to be written here.

Vertical size of the shadow can be found by introducing the function $f(r)$:
\begin{equation}
f(r) = y^2 = \sin^2 \Theta (1-\sin^2 \Psi) \, .
\end{equation}
Taking $df(r)/dr=0$ with $a=m$, we find $r_y(\vartheta_{\textrm{O}})$, and then we obtain the maximum value of vertical coordinate:
\begin{equation}
y_m(\vartheta_{\textrm{O}}) = \sqrt{f(r_y)} \, .
\end{equation}
Vertical size of the shadow is:
\begin{equation}
\Delta y = 2 \, y_m(\vartheta_{\textrm{O}}) \, .
\end{equation}

We now find the deformation $k$ as:
\begin{equation}
k(\delta, \vartheta_{\textrm{O}}) = \frac{\Delta x}{\Delta y} = \frac{F_0(\vartheta_{\textrm{O}}) + F_1(\vartheta_{\textrm{O}}) \sqrt{\delta}}{2y(\vartheta_{\textrm{O}})} \, .
\end{equation}

Supposing that the observer directly measures the value of $k$ and knows the angle $\vartheta_{\textrm{O}}$, we can write that:
\begin{equation}
\delta = \left( \frac{2y(\vartheta_{\textrm{O}}) k - F_0(\vartheta_{\textrm{O}})}{F_1(\vartheta_{\textrm{O}})} \right)^2 \, .
\end{equation}\\

\section{Calculation of BH spin for nearly equatorial observer}
\label{sec: nearly}

Our purpose is to get compact formulas, hence further we will consider the observer which is close to the equatorial plane ($\pi/2 - \vartheta_{\textrm{O}} \ll 1$). We will use the viewing angle $i$ instead the inclination angle $\vartheta_{\textrm{O}}$: $i=\pi/2 - \vartheta_{\textrm{O}}$. The angle $i$ indicates the inclination of the observer with respect to the equatorial plane; for observer in the equatorial plane $i=0$, for the polar observer $i=\pi/2$. We write in all formulas
\begin{equation}
\sin \vartheta_{\textrm{O}} = \sin (\pi/2 - i) = \cos i = 1 - \frac{i^2}{2} + \frac{i^4}{24} + ...
\end{equation}
and keep the small terms $\propto i^2$. We obtain:
\begin{equation}
x_L (\delta, i)= \left( - m - \frac{3}{2}m \, i^2 - \sqrt{6} m \sqrt{\delta} \right) / r_{\textrm{O}} \, ,
\end{equation} 
\begin{equation}
x_R (\delta, i) = \left( 8m - \frac{3}{2}m i^2 \right) / r_{\textrm{O}}  \, .
\end{equation} 
Horizontal size of the shadow is:
\begin{equation}
\Delta x = x_R- x_L = \left( 9m + \sqrt{6} m \sqrt{\delta}  \right) / r_{\textrm{O}} \, .
\end{equation}
We see that up to the terms proportional to $i^2$, the horizontal diameter of the BH does not depend on the viewing angle: if the observer looks at the extreme Kerr BH shadow and rises over the equatorial plane, the shadow is shifted to the 'left' as a whole. At the same time, the vertical diameter is becoming smaller:
\begin{equation}
\Delta y = \left(  6 \sqrt{3} m - \frac{\sqrt{3}}{3} m i^2 \right) / r_{\textrm{O}} .
\end{equation}

For oblateness $k = \Delta x / \Delta y$ we obtain:
\begin{equation}
k (\delta, i) = \frac{\sqrt{3}}{2} + \frac{\sqrt{18}}{18} \sqrt{\delta} + \frac{\sqrt{3}}{36} \, i^2 + \frac{\sqrt{18}}{324} \, i^2 \sqrt{\delta} \, .
\end{equation}
And expression of the spin via oblateness and the viewing angle is:
\begin{equation} \label{main-result2}
\delta =  18 \left( k - \frac{\sqrt{3}}{2} \right)^2 - 2 k \left( k - \frac{\sqrt{3}}{2} \right) i^2 \, .
\end{equation}

For observer in the equatorial plane ($\vartheta_{\textrm{O}} = \pi/2$, $i=0$), we have:
\begin{equation}
k(\delta) = \frac{\sqrt{3}}{2} + \frac{\sqrt{18}}{18} \sqrt{\delta} \, ,
\end{equation}
and the BH spin is calculated as
\begin{equation} \label{main-result}
a=(1-\delta)m \, , \quad \delta = 18  \left( k - \frac{\sqrt{3}}{2} \right)^2 \, , \quad k = \frac{\Delta x}{\Delta y} \, .
\end{equation}
Value of $a$ calculated for the equatorial plane is the lower limit of the spin of the BH at a given oblateness $k$: if the observer is not located in the equatorial plane, the larger value of the spin is required to obtain the same deformation.

By the way, the particular case of the equatorial observer can be easily checked in frame of Bardeen's approach with using the photon impact parameters. It is known that for observer in equatorial plane the impact parameters of left and right borders depends on spin $a$ as \cite{Dymnikova1986}
\begin{equation}
b_1 = a+ 8 \cos^3 \left[ (\pi - \arccos(a))/3  \right] \, ,
\end{equation}
\begin{equation}
b_2 = a - 8 \cos^3 \left[ (\arccos(a))/3 \right] \, .
\end{equation}
Taking into account that vertical diameter equals to $6\sqrt{3}m$, and expanding with $\delta$, we reproduce formula (\ref{main-result}).

\begin{figure}
	\centerline{\hbox{\includegraphics[width=0.48\textwidth]{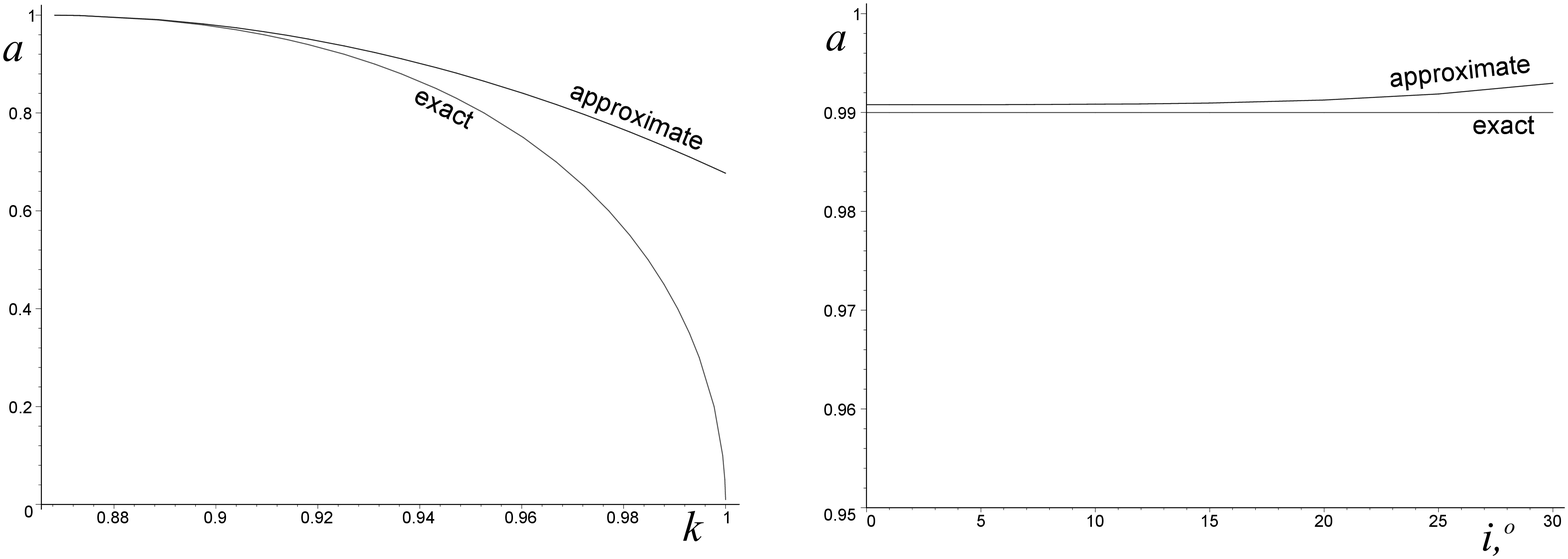}}}
	\caption{Accuracy of determination of black hole spin by formula (\ref{main-result2}) and (\ref{main-result}) in comparison with exact calculation, see details in the text.}
	\label{fig-accuracy}
\end{figure}

Let us now discuss in which range of parameters the resulting formula can be applied. At FIG. \ref{fig-accuracy}, LEFT, we present the calculation of spin $a$ at given oblateness $k$ for the equatorial plane. Horizontal axis shows all possible values of $k$. One curve represent the exact relation between $a$ and $k$ which is obtained by construction of number of shadow curves for different $a$ and evaluating their oblatenesses. Another curve shows the dependence of $a$ on $k$ according to (\ref{main-result}). It can be seen that for large values of $a$ the difference between two curves is very small. Speaking about possible range of $k$, we get that for the first half of the possible range of oblateness ($k \simeq 0.866 \div 0.933$, it corresponds to $\delta \simeq 0 \div 0.11$) the error on the spin $a$, at given the oblateness $k$, between the approximate analytic result and the exact result does not exceed $3\%$.

At FIG. \ref{fig-accuracy}, RIGHT, we present the exact and approximate calculation of the spin $a$ at given viewing angle $i$ for BH with $a=0.99m$. The exact curve is just $a=0.99$. To plot the approximate curve for this spin value, we take some value of $i$, construct the entire curve of the shadow, find the oblateness, and then obtain the approximate value of the spin with use of formula (\ref{main-result2}) with these known values of $k$ and $i$. Formula (\ref{main-result2}) contains $i^2$-terms, therefore the difference between analytical and exact results is very small even for relatively large values of $i$. For $ i=30 ^{\circ} $ the error on the spin $a$ does not exceed $0.3\%$.

\section{Conclusions}
Our conclusions:

(i) In the approximation of the nearly extreme Kerr BH with the spin $a=(1-\delta)m$ and a nearly equatorial observer with small $i$ (actually, relatively large values of $i$ would
be possible, see discussion above), we have investigated the dependence of the deformation on both the BH spin and the observer's viewing angle. Remarkably, we obtain that the dependence of the deformation on the spin is strong: the deformation is proportional to $\sqrt{\delta}$. It means that a small difference of the spin from $m$ can lead to a notable deviation of the observed deformation from the extreme value $\sqrt{3}/2$, see Fig. 3.

(ii) Knowing the oblateness by measuring the horizontal and vertical diameters of the shadow, one can easily obtain the lower limit on the BH spin by the formula (\ref{main-result}), without need to construct or model the entire curve of the shadow.

(iii) If the viewing angle is known from other observations, one can directly calculate the spin using (\ref{main-result2}).

(iv) In all situations when the shadow curve is noticeably different from the circular shape, our approximate formulas provide a high accuracy of calculation.

\section*{Acknowledgments}

This work is financially supported by Russian Science Foundation, Grant No. 15-12-30016. The author is thankful to anonymous referees for statement of some important points. The author would like to thank Volker Perlick for the introduction to this topic and useful discussions. The author would like to thank N. S. Voronova for the help with brushing up the language
of the present paper. The author brings special thanks to G.S. Bisnovatyi-Kogan for motivation, discussions and permanent support of all scientific initiatives.

\bibliographystyle{ieeetr}

\end{document}